\documentclass[iop,apj]{emulateapj}
\usepackage{url,xspace}
\usepackage[usenames, dvipsnames]{color}
\usepackage[breaklinks,colorlinks,citecolor=blue]{hyperref}
\shorttitle{SAMI stellar population drivers}
\shortauthors{Barone et al.}

\begin{document}

\title{The SAMI Galaxy Survey: gravitational potential and surface density drive stellar populations -- I. early-type galaxies}

\author{Tania M. Barone\altaffilmark{1,2}, 
Francesco D'Eugenio\altaffilmark{1,2},
Matthew Colless\altaffilmark{1,2},
Nicholas Scott\altaffilmark{2,3},
Jesse van de Sande\altaffilmark{3},
Joss Bland-Hawthorn\altaffilmark{2,3},
Sarah Brough\altaffilmark{4},
Julia J. Bryant\altaffilmark{2,3,5},
Luca Cortese\altaffilmark{6},
Scott M. Croom\altaffilmark{2,3},
Caroline Foster\altaffilmark{3},
Michael Goodwin\altaffilmark{5},
Iraklis S. Konstantopoulos\altaffilmark{7},
Jon S. Lawrence\altaffilmark{5},
Nuria P. F. Lorente\altaffilmark{5},
Anne M. Medling\altaffilmark{1,8,9} 
Matt S. Owers\altaffilmark{5,10},
Samuel N. Richards\altaffilmark{11}}

\affil{$^{1}$ Research School of Astronomy \& Astrophysics, The Australian National University, Cotter Road, Weston Creek, ACT 2611, Australia\\
$^{2}$ ARC Centre of Excellence for All-sky Astrophysics (CAASTRO)\\
$^{3}$ Sydney Institute for Astronomy, School of Physics, A28, The University of Sydney, NSW, 2006, Australia\\
$^{4}$ School of Physics, University of New South Wales, NSW 2052, Australia\\
$^{5}$ Australian Astronomical Observatory, 105 Delhi Rd, North Ryde, NSW 2113\\
$^{6}$ International Centre for Radio Astronomy Research, University of Western Australia, 35 Stirling Highway, Crawley WA 6009, Australia\\
$^{7}$ Atlassian 341 George St Sydney, NSW 2000\\
$^{8}$ Cahill Center for Astronomy and Astrophysics California Institute of Technology, MS 249-17 Pasadena, CA 91125, USA \\
$^{9}$ Hubble Fellow \\
$^{10}$ Department of Physics and Astronomy, Macquarie University, NSW 2109\\
$^{11}$ SOFIA Operations Center, USRA, NASA Armstrong Flight Research Center, 2825 East Avenue P, Palmdale, CA 93550, USA
}

\begin{abstract}
The well-established correlations between the mass of a galaxy and the properties of its stars are considered evidence for mass driving the evolution of the stellar population. However, for early-type galaxies (ETGs), we find that $g-i$ color and stellar metallicity [Z/H] correlate more strongly with gravitational potential $\Phi$ than with mass $M$, whereas stellar population age correlates best with surface density $\Sigma$. Specifically, for our sample of 625 ETGs with integral-field spectroscopy from the SAMI Galaxy Survey, compared to correlations with mass, the color--$\Phi$, [Z/H]--$\Phi$, and age--$\Sigma$ relations show both smaller scatter and less residual trend with galaxy size. For the star formation duration proxy [$\alpha$/Fe], we find comparable results for trends with $\Phi$ and $\Sigma$, with both being significantly stronger than the [$\alpha$/Fe]-$M$ relation. In determining the strength of a trend, we analyze both the overall scatter, and the observational uncertainty on the parameters, in order to compare the intrinsic scatter in each correlation. These results lead us to the following inferences and interpretations: (1) the color--$\Phi$ diagram is a more precise tool for determining the developmental stage of the stellar population than the conventional color--mass diagram; and (2) gravitational potential is the primary regulator of global stellar metallicity, via its relation to the gas escape velocity. Furthermore, we propose the following two mechanisms for the age and [$\alpha$/Fe] relations with $\Sigma$: (a) the age--$\Sigma$ and [$\alpha$/Fe]--$\Sigma$ correlations arise as results of compactness driven quenching mechanisms; and/or (b) as fossil records of the $\Sigma_{SFR}\propto\Sigma_{gas}$ relation in their disk-dominated progenitors.
\end{abstract}

\keywords{galaxies: evolution --- galaxies: fundamental parameters --- galaxies: kinematics and dynamics}

\section{Introduction}

Studying the stellar population (SP) of a galaxy is key to understanding its formation and evolution. By using different parameters, we can piece together various aspects of the galaxy's history. Photometric colors provide a robust, directly observable parameter for analyzing SPs (e.g. \citealt{Tinsley1980}). However, many SP parameters appear degenerate in optical photometry; for example age, metallicity, and reddening due to dust extinction. This restricts the accuracy of SP analyses using colors. Early spectroscopic observations identified spectral features which have varying dependencies on these parameters, allowing us to break the apparent degeneracy and obtain well constrained SP parameters \citep{WortheyG1994}. One popular method is the Lick indices system, that uses the strength of specific optical absorption lines to quantify galaxy SPs \citep{Worthey1994}. SP properties such as age, [Z/H] and [$\alpha/Fe$] are then obtained by comparing values of specific Lick indices with SP models.

The well known SP--stellar mass correlation is often considered evidence of stellar mass driving SP evolution \citep[e.g.][]{Gallazzi2005,Peng2010,Dave2011}. Even so, SP parameters correlate with several other galaxy properties including velocity dispersion, large-scale environment, and surface brightness, making it unclear which correlations are causal and which are the result of another underlying trend \citep{Thomas2005,Nelan2005,Sanchez-Blazquez2006,Smith2007,Franx2008,Graves2009_1,Graves2009_2,Wake2012,McDermid2015}. Without understanding the observational uncertainty on these parameters, we cannot know the intrinsic scatter, and hence which relations are fundamentally tighter. Additionally, many SP analyses have relied on single-fiber spectroscopy, which is subject to aperture bias (e.g.  6dFGS; \citealt{Jones2004}, SDSS; \citealt{York2000}, GAMA; \citealt{Driver2011}). Radial trends within galaxies combined with aperture bias can produce spurious global trends; for example, the radial metallicity trend within ETGs can appear as a trend between global [Z/H] and size.

More recent surveys instead use integral-field spectroscopy, sampling the light across most of the galaxy and so mitigating aperture effects (e.g. SAURON: \citealt{deZeeuw2002}; ATLAS3D: \citealt{Cappellari2011_I}; CALIFA: \citealt{Sanchez2012}; MaNGA: \citealt{Bundy2015}). We use data from the SAMI Galaxy Survey \citep{Bryant2015}, an integral-field survey using the Sydney-AAO Multi-object Integral-field spectrograph \citep[SAMI;][]{Croom2012}. This Paper is followed by a companion paper D'Eugenio et al. (2018 in prep., hereafter Paper II). Here, our analysis focuses on the SPs of morphologically selected ETGs from SAMI; Paper II focuses on constraining color relations using color-selected samples from the Galaxy And Mass Assembly survey \citep[GAMA;][]{Driver2011} as well as SAMI. Our aim is to build on recent studies examining SP trends with aperture velocity dispersion $\sigma$ \citep{Graves2009_1,Thomas2010,Wake2012} and surface density $\Sigma$ \citep{Scott2017}. We want to understand which relations have the lowest \emph{intrinsic} scatter, in order to distinguish between fundamental correlations, and what is the result of some other underlying trend. However, the absolute intrinsic scatter is difficult to measure because it depends strongly on the assumed measurement uncertainties. Instead, we can use the necessary condition that, due to the non-zero uncertainty on radius $R$, $M R^x$ must have a higher observational uncertainty than $M$ alone (for $x\neq 0$). Using this principle, and comparing the observed scatter about the fits, we can rank the relations based on their \textit{relative} intrinsic scatter. With this approach, we study the correlations between SP and galaxy structural parameters, specifically mass $M$, gravitational potential $\Phi\propto M/R$, and surface density $\Sigma\propto M/R^2$. For each structural parameter we define two estimators, one based on spectroscopic velocity dispersion (henceforth called the \textit{spectroscopic} estimators), the other based on photometric stellar masses (the \textit{photometric} estimators). Within each set of estimators (i.e. the spectroscopic or photometric), the three structural parameters differ only by factors of the effective radius, allowing us to directly compare the observational uncertainty and hence infer the relative intrinsic scatter in the relations. We also look at the residuals of each trend with galaxy size. With this robust analysis, we aim to determine the primary physical factors determining galaxy SPs, and the mechanisms which drive their evolution. Throughout this Paper we assume a $\Lambda$CDM universe with $\Omega_m=0.3$, $\Omega_{\lambda}=0.7$, and $H_0=70$ km/s/Mpc.

\section{The SAMI Galaxy Survey} 

The SAMI Galaxy Survey is a presently ongoing, integral-field survey aiming to observe up to 3600 galaxies by the end of 2018. The survey uses the SAMI instrument installed on the 3.9m Anglo-Australian Telescope, connected to the AAOmega spectrograph (\citealt{Sharp2006}; see \citealt{Sharp2015} for data reduction). The sample is mass selected, however the mass limit varies depending upon the redshift range. Details of the target selection and input catalogs are described in \cite{Bryant2015}, with the cluster galaxies further described in \cite{Owers2017}. The SAMI spectrograph uses 13 fused-fibre hexabundles \citep{Bland-Hawthorn2011,Bryant2014}, each composed of 61 individual fibres, tightly packed to form an approximately circular grid 15 arcsec  in diameter. We use data from internal release v0.9.1, comprising 1380 galaxies with low redshifts ($z<0.1$) and a broad range of stellar masses $10^7<M_*<10^{12}$ (\citealt{Allen2015}; see \citealt{Green2018} for data release 1). We define a subsample of 625 ETGs having a visual morphological classification of elliptical, lenticular, or early spiral \citep{Cortese2016}. Excluding early spirals from our sample does not change our conclusions.

We experimented with different samples, including a mass-function weighted sample using weights based on the stellar mass function of \cite{Kelvin2014}, which gives the effective number of galaxies per unit volume in a stellar mass interval. The weights were calculated by taking the ratio between the stellar mass function, and the actual number of observed SAMI galaxies in each stellar mass interval. The results of this analysis are summarized in Table \ref{tbl}, alongside the results of the analysis without weights. We find consistent results between the original SAMI sample (which is mass-limited in redshift bins) and the mass-function weighted sample (which approximates a sample with a single mass limit). Since the two analyses are consistent, to avoid over-dependence on this theoretical model, we focus our analysis on the results without weights.

We use $g-i$ color as a simple, directly observable parameter for comparing SPs; we use the dust-uncorrected values to remain model-independent. For the ETG subsample, we use the single-burst equivalent, luminosity-weighted SP parameters age, metallicity [Z/H], and $\alpha$-element abundance [$\alpha$/Fe] from \cite{Scott2017}. Stellar masses, $M_*$, were obtained from $g-i$ color by \cite{Bryant2015} and \cite{Owers2017} following the method of \cite{Taylor2011}:

\begin{equation}
\log_{10}\frac{M_*}{M_{\odot}}=1.15+0.70(g-i)_{rest}-0.4 M_i
\end{equation}

Where $M_i$ is the rest frame i-band absolute AB magnitude, and $M_*$ has solar mass units.

Effective radii ($R_e$) were measured using Multi-Gaussian Expansion modeling \citep{Cappellari2002} from r-band images (Paper II); $R_e$ is the projected, circularised radius enclosing half the total light. The luminosity-weighted, line-of-sight velocity dispersion ($\sigma$) within 1$R_e$ was then measured as in \cite{vandeSande2017}.

\begin{table*}
\begin{tabular}{cc | cccc | cccc} \hline\hline \multicolumn{2}{}{} & \multicolumn{4}{|c|}{Unweighted} & \multicolumn{4}{|c}{Mass-function Weighted}\\
Y-axis & X-axis & $\mathrm{RMS_G}$ & $\mathrm{\rho_S}$ & $\frac{a_r}{\Delta a_r}$ & $\mathrm{RMS_{rm}}$ & $\mathrm{RMS_{G}}$ & $\rho_S$ & $\frac{a_{r}}{\Delta a_r}$ & $\mathrm{RMS}_{\mathrm{rm}}$ \\ \hline
$g-i_{all}$ & $M_*$ & $0.1589\pm \;0.0004$ & 0.78 & 19.7 & 0.1586 & $0.1589\pm \;0.0004$ & 0.78 & 19.8 & 0.1586 \\
$g-i_{all}$ & $M_*/R_e$ & $0.1269\pm \;0.0005$ & 0.87 & 9.0 & 0.1277 & $0.1269\pm \;0.0005$ & 0.87 & 7.1 & 0.1277 \\
$g-i_{all}$ & $M_*/R_e^2$ & $0.1438\pm \;0.0008$ & 0.82 & -13.9 & 0.1370 & $0.1438\pm \;0.0008$ & 0.82 & -16.4 & 0.1370 \\
$g-i_{ETG}$ & $M_*$ & $0.0910\pm \;0.0004$ & 0.50 & 7.3 & 0.0896 & $0.0962\pm \;0.0004$ & 0.50 & 10.0 & 0.0954 \\
$g-i_{ETG}$ & $M_*/R_e$ & $0.0816\pm \;0.0012$ & 0.67 & 2.7 & 0.0786 & $0.0839\pm \;0.0014$ & 0.67 & 5.2 & 0.0797 \\
$g-i_{ETG}$ & $M_*/R_e^2$ & $0.0929\pm \;0.0014$ & 0.44 & -11.5 & 0.0891 & $0.0961\pm \;0.0015$ & 0.45 & -9.6 & 0.0918 \\
\rule{0pt}{4ex}{[Z/H]} & $M_D$ & $0.1678\pm \;0.0003$ & 0.38 & 4.7 & 0.1664 & $0.1866\pm \;0.0004$ & 0.37 & 5.0 & 0.1855 \\
{[Z/H]} & $M_D/R_e$ & $0.1534\pm \;0.0002$ & 0.47 & 3.6 & 0.1531 & $0.1708\pm \;0.0006$ & 0.50 & 2.6 & 0.1719 \\
{[Z/H]} & $M_D/R_e^2$ & $0.1750\pm \;0.0002$ & 0.44 & -6.2 & 0.1738 & $0.1834\pm \;0.0005$ & 0.43 & -5.3 & 0.1829 \\
{[Z/H]} & $M_*$ & $0.1647\pm \;0.0003$ & 0.40 & 7.5 & 0.1647 & $0.1773\pm \;0.0003$ & 0.41 & 6.7 & 0.1776 \\
{[Z/H]} & $M_*/R_e$ & $0.1549\pm \;0.0015$ & 0.50 & 3.0 & 0.1515 & $0.1652\pm \;0.0013$ & 0.53 & 1.0 & 0.1648 \\
{[Z/H]} & $M_*/R_e^2$ & $0.1766\pm \;0.0012$ & 0.41 & -11.2 & 0.1736 & $0.1876\pm \;0.0009$ & 0.39 & -10.0 & 0.1855 \\
\rule{0pt}{4ex}Age & $M_D$ & $0.2283\pm \;0.0024$ & 0.15 & 10.3 & 0.2275 & $0.2281\pm \;0.0024$ & 0.17 & 8.7 & 0.2264 \\
Age & $M_D/R_e$ & $0.2183\pm \;0.0045$ & 0.42 & 10.0 & 0.2095 & $0.2137\pm \;0.0046$ & 0.47 & 9.0 & 0.2053 \\
Age & $M_D/R_e^2$ & $0.1993\pm \;0.0041$ & 0.57 & -0.6 & 0.1911 & $0.1954\pm \;0.0044$ & 0.60 & -0.5 & 0.1867 \\
Age & $M_*$ & $0.2330\pm \;0.0045$ & 0.10 & 14.9 & 0.2253 & $0.2370\pm \;0.0045$ & 0.14 & 8.3 & 0.2312 \\
Age & $M_*/R_e$ & $0.2484\pm \;0.0113$ & 0.39 & 7.4 & 0.2045 & $0.2391\pm \;0.0106$ & 0.45 & 5.8 & 0.2019 \\
Age & $M_*/R_e^2$ & $0.2389\pm \;0.0098$ & 0.47 & -7.1 & 0.2047 & $0.2348\pm \;0.0102$ & 0.47 & -6.0 & 0.2029 \\
\rule{0pt}{4ex}{[$\alpha$/Fe]} & $M_D$ & $0.1049\pm \;0.0007$ & 0.32 & 7.8 & 0.1046 & $0.1072\pm \;0.0006$ & 0.27 & 4.9 & 0.1075 \\
{[$\alpha$/Fe]} & $M_D/R_e$ & $0.0961\pm \;0.0007$ & 0.49 & 3.6 & 0.0946 & $0.0999\pm \;0.0009$ & 0.43 & 4.0 & 0.0982 \\
{[$\alpha$/Fe]} & $M_D/R_e^2$ & $0.0963\pm \;0.0007$ & 0.51 & -5.4 & 0.0954 & $0.1009\pm \;0.0010$ & 0.45 & -3.6 & 0.0983 \\
{[$\alpha$/Fe]} & $M_*$ & $0.1094\pm \;0.0007$ & 0.21 & 2.1 & 0.1084 & $0.1110\pm \;0.0007$ & 0.20 & 2.9 & 0.1103 \\
{[$\alpha$/Fe]} & $M_*/R_e$ & $0.1032\pm \;0.0015$ & 0.39 & 1.0 & 0.1019 & $0.1073\pm \;0.0018$ & 0.34 & 0.8 & 0.1056 \\
{[$\alpha$/Fe]} & $M_*/R_e^2$ & $0.1076\pm \;0.0014$ & 0.30 & -6.7 & 0.1066 & $0.1099\pm \;0.0013$ & 0.23 & -5.2 & 0.1087 \\
\rule[-0.5ex]{0pt}{0pt} \\ \hline\hline \end{tabular}  \centering \caption{Summary of the results for both the unweighted, and the mass-function weighted analyses. $\mathrm{RMS_G}$ and $\mathrm{RMS_{rm}}$ indicate the RMS values about the Gaussian model fit, and the running median respectively. $\rho_S$ represents the Spearman correlation coefficient. $\frac{a_r}{\Delta a_r}$ shows the $\sigma$ significance of the residual trend with size, where $a_r$ is the slope of the residual trend with 1$\sigma$ uncertainty $\Delta a_r$} \label{tbl}
\end{table*}

We define spectroscopic estimators for the gravitational potential $\Phi\propto\sigma^2$ and surface density $\Sigma\propto\sigma^2/R_e$ by assuming galaxies are structurally homologous and in virial equilibrium. We use the virial theorem to also define the spectroscopic (dynamical) mass proxy $M_D\equiv\sigma^{2}R_e/(3G)$ (the arbitrary 1/3 scaling factor conveniently makes $M_D$ span the same range as $M_*$). Further assuming a uniform dark matter fraction within $1R_e$, we define the photometric estimators $\Phi\propto M_*/R_e$ and $\Sigma\propto M_*/R_e^2$. Hence we have two independent methods for estimating mass, gravitational potential, and surface density: $M_*$, $M_*/R_e$ and $M_*/R_e^2$ rely solely on photometry, whereas $M_D$, $M_D/R_e$ and $M_D/R_e^2$ also use spectroscopy. In the limit that galaxies are virialized and have the same mass-to-light ratio, these measures would be proportional. See Paper II for a comparison of $M_*$ and $M_D$. We note that $M_*$ is calculated under the assumption of a uniform \cite{Chabrier2003} initial mass function (IMF). However, the IMF may vary systematically with stellar mass-to-light ratio, leading to an underestimated $M_*$ for massive galaxies \citep{Cappellari2012}. Despite this bias, the photometric results are remarkably consistent with the spectroscopic results, and are included to provide an independent measure for each structural parameter with uncorrelated uncertainties. In addition, photometric observations are significantly less expensive than spectroscopy.

\section{Methods and Results}

\begin{figure*}
\includegraphics[type=pdf, ext=.pdf, read=.pdf, width=1.0\textwidth]{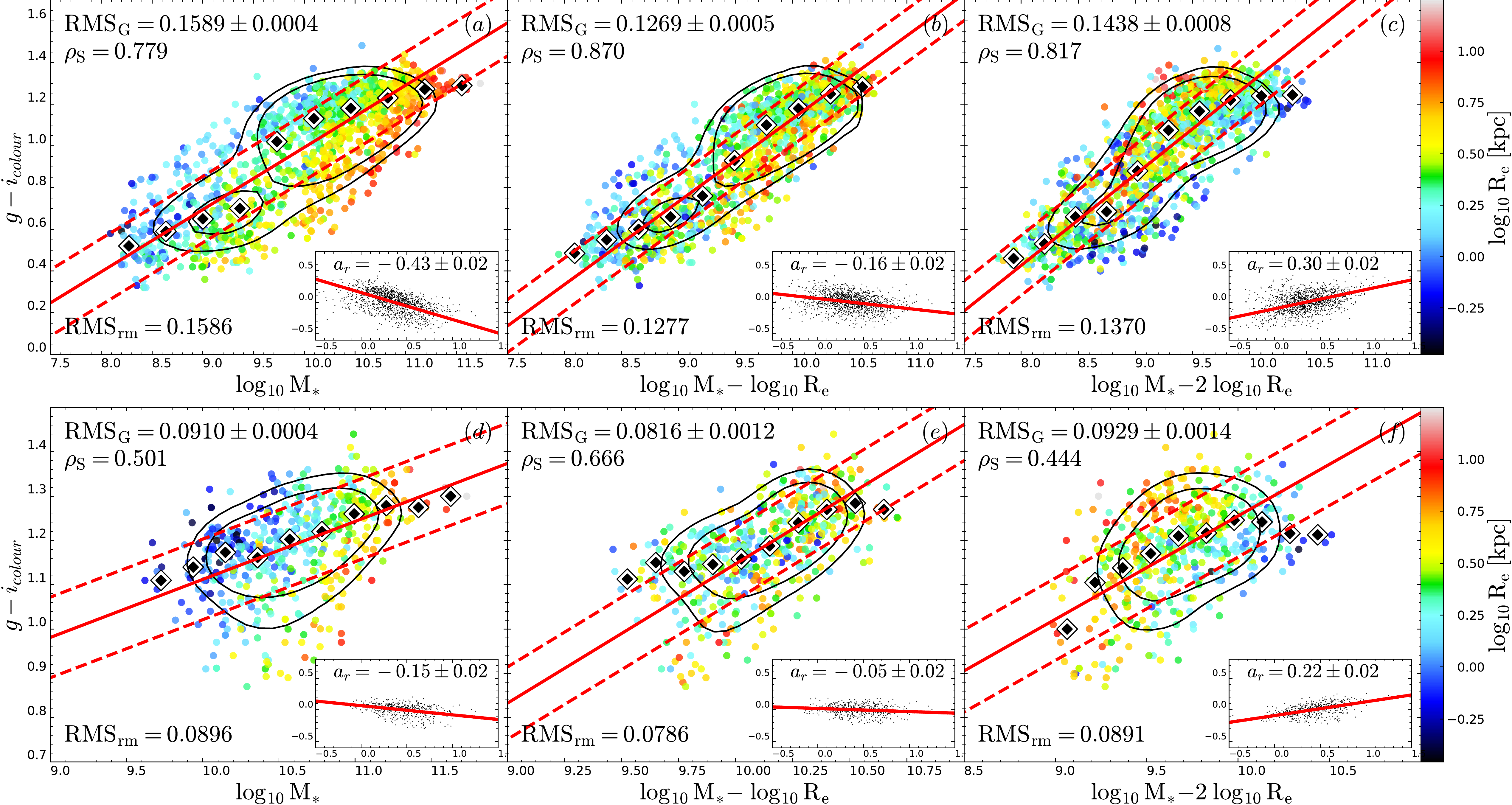} \centering
\caption{$g-i$ color versus $M_*$, $M_*/R_e$, and $M_*/R_e^2$ for the full sample (top row) and for the ETG subsample (bottom row). The solid red line is the best-fit linear relation and the dashed red lines indicate the RMS about this fit. The RMS of the best-fit line ($\mathrm{RMS}_{G}$) with its 1$\sigma$ uncertainty is given at the top of each panel, along with the Spearman coefficient $\rho_{S}$. The black diamonds show the running median in evenly spaced bins, and the RMS about this running median ($\mathrm{RMS}_{rm}$) is shown in the bottom left of the panels. The contours enclose $60\%$ and $80\%$ of the data. The color scale indicates $R_e$ in units of $\log(\mathrm{kpc})$. The inset panels show the best-fit residuals as a function of $\log R_e$. The slope of the residual trend $a_r$ is displayed at the top of each inset. For both the full SAMI sample and the ETG subsample, the color--$M_*/R_e$ relations (panels b and e) have less scatter (lower $\mathrm{RMS}_{G}$ and $\mathrm{RMS}_{rm}$), are more significant (higher $\rho_S$), and have less residual trend with radius (demonstrated by the inset panels) compared to the relations with $M_*$ or $M_*/R_e^2$.}\label{gi}
\end{figure*}

We fit linear relations via a maximum likelihood optimization followed by Markov Chain Monte Carlo (MCMC) integration \citep{GoodmanWeare2010}. The data is modeled as a two-dimensional Gaussian, which avoids bias inherent to orthogonal or parallel least squares regressions \citep[see e.g.][]{Magoulas2012}. The log-likelihood function is optimized using the method of Differential Evolution \citep{Storn1997}. For all the relations except for age, we perform outlier rejection by omitting points which lie outside the 90\% contour line. Due to the larger scatter, in the age relations we perform the outlier rejection at the 80\% contour. We calculate the root-mean-square about the Gaussian model fit ($\mathrm{RMS}_{G}$), which is displayed at the top left on each panel.

In order to assess whether the linear fit is an accurate model, we compute a running median using equally sized bins in log-space. For all the correlations that we consider to be physically motivated, the running median closely follows the log-linear fits, supporting our choice of model. The RMS about the running median ($\mathrm{RMS}_{rm}$) is shown at the bottom left in the panels.

For each relation, we also fit the residuals about the Gaussian model as a function of $R_e$, using the same method as for the main relation. These residual fits indicate which of $M$, $\Phi$, or $\Sigma$ best encapsulates the SP parameter's dependence on galaxy size. The errors from the initial fit are incorporated into the uncertainty on the residual values, which in turn is taken into account when fitting the residuals.

We use $\mathrm{RMS}_{G}$ and $\mathrm{RMS_{rm}}$ to determine the quality of the relation, and the Spearman coefficient ($\rho_S$) to define the significance of the trend. We estimate the uncertainties on each parameter by full integration of the posterior distribution. Our results remain unchanged whether we use the median-absolute-deviation (MAD) or RMS.

Due to the relatively small sample size, plane fits of SP parameters as log-linear combinations of $M_D$ (or $M_*$) and $R_e$ were poorly constrained, and hence omitted.

We firstly compare how $g-i$ color trends with the photometric estimators $M_*$, $M_*/R_e$ and $M_*/R_e^2$ using both the full sample and the ETG subsample. Although $M_*$ has an explicit dependence on $g-i$ color, we also use $M_*$ to estimate all three proxies, so any bias due to this explicit dependence will not affect the comparison. For an analysis using spectral energy distribution masses, see Paper II. We can rule out a correlation in the uncertainties due to random errors on $M_*$ and $R_e$, because $R_e$ uses $r$-band photometry whereas $M_*$ uses $g$- and $i$-band magnitudes. We then use the ETG subsample to fit [Z/H], age, and [$\alpha$/Fe] as functions of $M$, $\Phi$, and $\Sigma$ using both the spectroscopic and photometric measures.

We perform an identical analysis on the mass-function weighted sample, and summarize the results in Table \ref{tbl}. Given the analyses show consistent results, in this section we focus on the unweighted analysis.

\subsection{$g-i$ Color}
Figure \ref{gi}a shows $g-i$ color as a function of $M_*$ for the full sample, and exhibits the well-documented bimodal trend of color--mass diagrams, with galaxies forming a red sequence (RS) and blue cloud (BC). As the contour lines reveal, the RS and BC do not align in color--$M_*$ space, and so the best-fit line does not accurately model the distinct distributions; it simply provides a reference for comparison of the RS and BC alignment for the different relations. The color scale indicates galaxy size and shows a strong residual trend, implying that, at fixed mass, size contains additional information on $g-i$ color. By comparison, Figure \ref{gi}b shows that in the color--$M_*/R_e$ diagram the residual trend with size is significantly less; furthermore, the RS and BC are better aligned in Figure \ref{gi}b, as apparent from the contours and demonstrated by the lower $\mathrm{RMS_G}=0.127$ (cf.\ $\mathrm{RMS_G}=0.159$ for color--$M_*$). Similarly, comparing color as a function of $M_*/R_e$ (Figure \ref{gi}b) and of $M_*/R_e^2$ (Figure \ref{gi}c), the RMS values are smaller and the residual trend with size is less significant for color--$M_*/R_e$. 

The bottom row of Figure \ref{gi} shows these relations for the ETG subsample (effectively for the RS only). For ETGs, $g-i$ color has a stronger and tighter relation with $M_*/R_e$ compared to either $M_*$ or $M_*/R_e^2$, and less residual trend with size. By construction, $M_*/R_e$ necessarily has a larger observational uncertainty than $M_*$ alone, as $M_*/R_e$ includes the uncertainty on both $M_*$ and $R_e$. Yet the color--$M_*/R_e$ relation shows less scatter than color--$M_*$, therefore color--$M_*/R_e$ must have significantly lower \textit{intrinsic} scatter. Furthermore, color--$M_*/R_e$ has a higher Spearman coefficient of $\rho_S=0.666$, compared to $\rho_S=0.501$ and $0.444$ for $M_*$ and $M_*/R_e^2$ respectively. In Paper II we find similar results for the BC: color--$M_*/R_e$ has less scatter and less residual trend with size compared to color--$M_*$. For both the total sample and the ETG subsample, compared to trends with $M_*$ and $M_*/R_e^2$, the color--$M_*/R_e$ relation has the lowest RMS values, least residual trend with size, and highest $\rho_S$.

\begin{figure*}
\includegraphics[type=pdf, ext=.pdf, read=.pdf, width=1.0\textwidth]{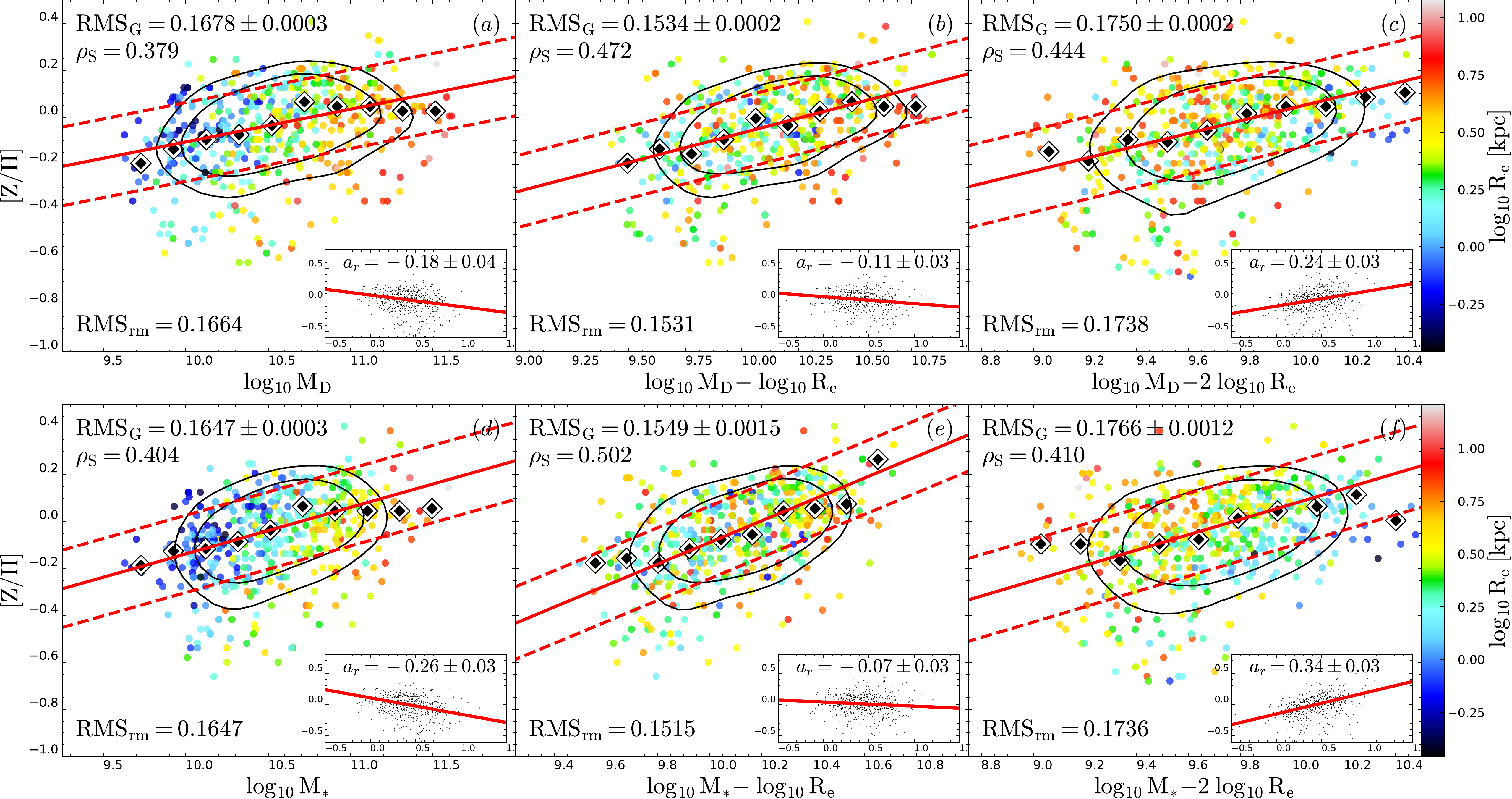} \centering
\caption{[Z/H] versus $M$, $\Phi$, and $\Sigma$ for ETGs. The top row uses the spectroscopic estimator $M_D\propto\sigma^{2}R_e$, the bottom row uses the purely photometric $M_*$. The inset panels show the best-fit residuals as a function of $\log R_e$ (other details as for Figure 1). For both the spectroscopic and photometric estimators, the [Z/H]--$\Phi$ relations (panels b and e) have the least scatter (lowest $\mathrm{RMS}_G$ and $\mathrm{RMS}_{rm}$), are the most significant (highest $\rho_S$), and have the least residual trend with radius (inset panel).}\label{met}
\end{figure*}

\begin{figure*}
\includegraphics[type=pdf, ext=.pdf, read=.pdf, width=1.0\textwidth]{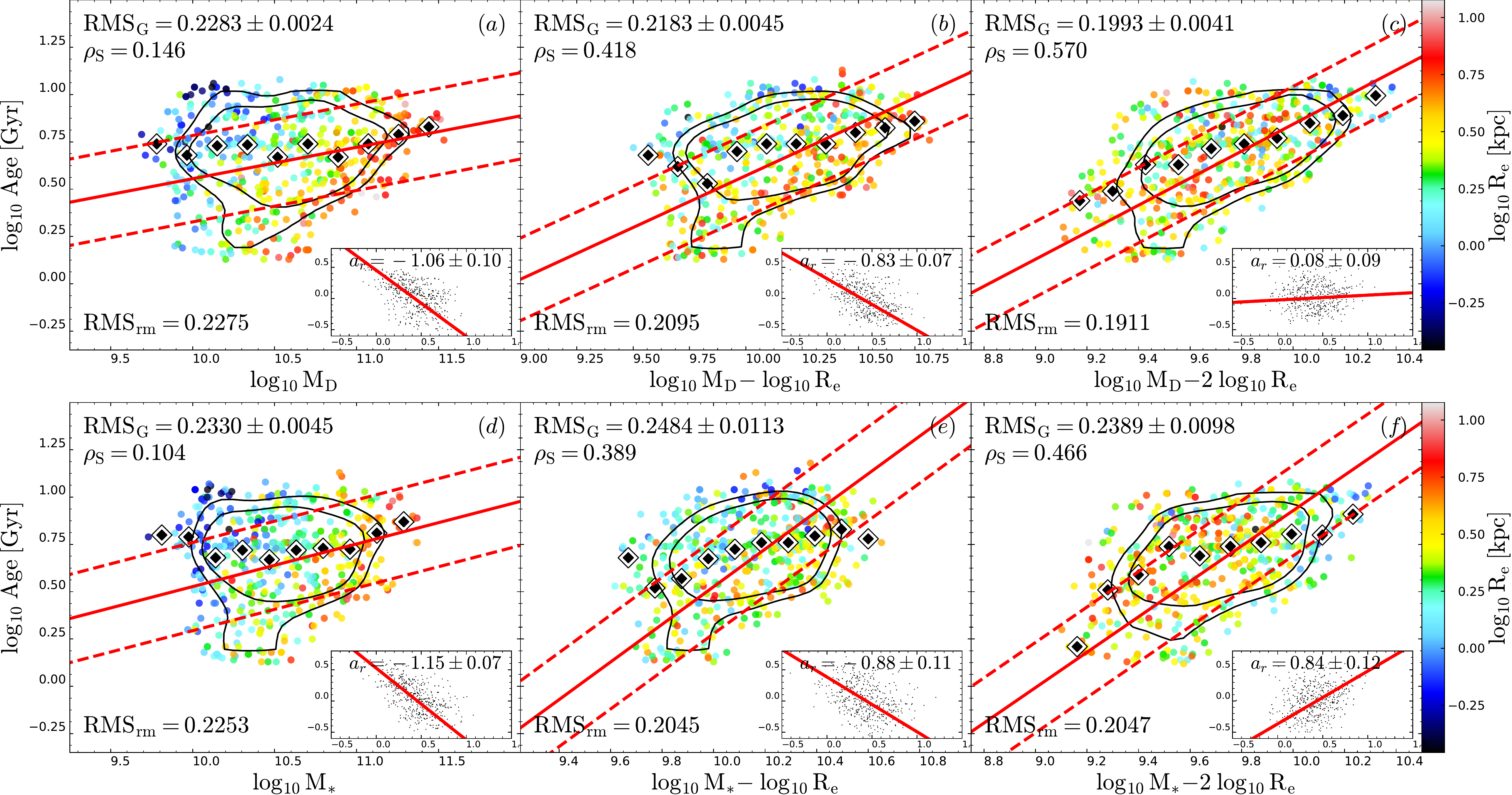} \centering
\caption{Age versus $M$, $\Phi$, and $\Sigma$ for ETGs. The top row uses the spectroscopic estimator $M_D\propto\sigma^{2}R_e$, the bottom row uses the purely photometric $M_*$. The inset panels show the best-fit residuals as a function of $\log R_e$ (other details as for Figure 1). Overall, for both the spectroscopic and photometric estimators, the age--$\Sigma$ relations (panels c and f) tend to have the least scatter (lowest $\mathrm{RMS}_G$ and $\mathrm{RMS}_{rm}$), are the most significant (highest $\rho_S$), and have the least residual trend with radius (inset panel).}\label{age}
\end{figure*}

\subsection{Metallicity}
In Figure \ref{met} we show the relations between [Z/H] and $M$, $\Phi$, and $\Sigma$; the top row uses spectroscopic virial masses and the bottom row photometric stellar masses. We see consistent results between the spectroscopic and photometric mass estimators. With increasing power of $R_e$, the residual trend with size goes from negative in the [Z/H]--$M$ relations, to close to zero for [Z/H]--$\Phi$, and finally to positive for [Z/H]--$\Sigma$. The [Z/H]--$\Phi$ relations also have the tightest and most significant correlations; [Z/H]--$M_*/R_e$ has an $\mathrm{RMS_G}$ = 0.155, whereas the $\mathrm{RMS_G}$ values for [Z/H]--$M_*$ and [Z/H]--$M_*/R^2_e$ are higher by $7\sigma$ and $14\sigma$ respectively.
Given the higher observational uncertainty on $M_*/R_e$ than $M_*$ alone, the lower RMS for [Z/H]--$M/R_e$ implies this relation must also have a lower intrinsic scatter than [Z/H]--$M$.
For the spectroscopic estimators, $M_D/R_e\propto\sigma^2$ and hence has a lower observational uncertainty than $M_D$ and $M_D/R_e^2$, and so we cannot comment on the relative intrinsic scatter about these trends. The result is, however, consistent with the photometric estimators, with [Z/H]--$M_D/R_e$ showing the lowest RMS. The two [Z/H]--$\Phi$ relations also show the highest $\rho_S$.

\subsection{Age}
We show the results of our analysis for age in Figure~\ref{age}. There is more scatter in the age relations than in the other SP parameters, most likely because age is more sensitive to recent bursts of star formation \citep{SerraTrager2007}. Despite this larger scatter, we see statistically significant results.

Age is well-known to have a dependence on galaxy mass \citep[e.g.][]{Kauffmann2003,Gallazzi2005,Thomas2010,McDermid2015}, however age--$M_D$ (Figure \ref{age}d) shows only a weak correlation, and a large residual trend with size. Age--$M_D$ also has a lower Spearman coefficient than $M_D/R_e$ and $M_D/R_e^2$. Focusing instead on $\Sigma$, we see that age--$M_D/R_e^2$ has the lowest $\mathrm{RMS_G}$ = 0.200, highest Spearman coefficient $\rho_S=0.570$, and a residual trend with size statistically consistent with zero (within 1$\sigma$). $M_D/R_e^2$ and $M_D$ have the same observational uncertainty, which is by construction greater than the uncertainty for $M_D/R_e$. The notably lower RMS for age--$M_D/R_e^2$ therefore implies the intrinsic scatter in this trend must also be significantly lower. We find consistent results for the photometric estimators; $M_*/R_e^2$ has the lowest intrinsic scatter and largest $\rho_S$. However there are large residual trends with size for all three photometric parameters, likely due to the large scatter in the age measurements.

\begin{figure*}
\includegraphics[type=pdf, ext=.pdf, read=.pdf, width=1.0\textwidth]{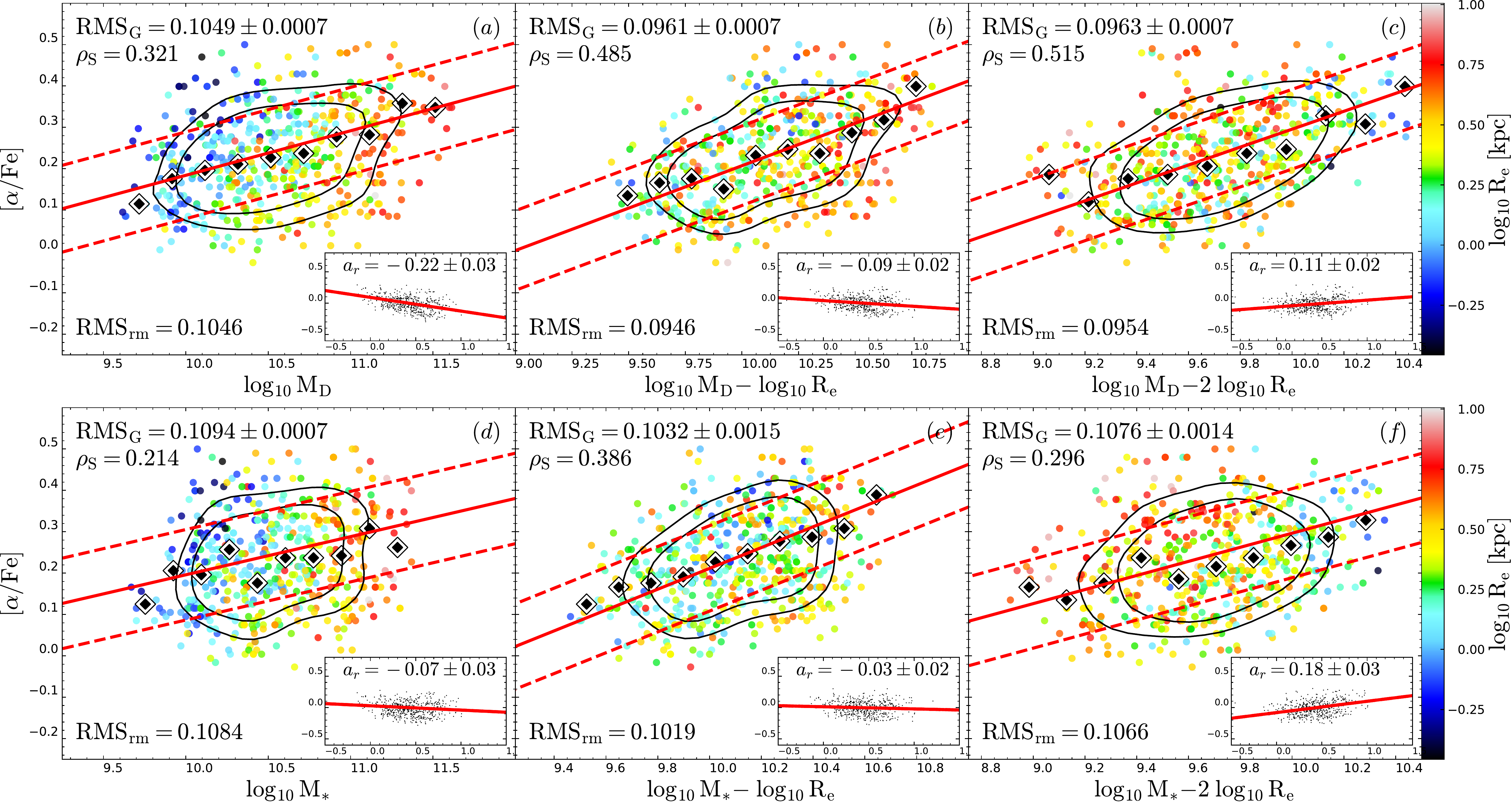} \centering
\caption{[$\alpha$/Fe] versus $M$, $\Phi$, and $\Sigma$ for ETGs. The top row uses the spectroscopic estimator $M_D\propto\sigma^{2}R_e$, the bottom row uses the purely photometric $M_*$. The inset panels show the best-fit residuals as a function of $\log R_e$ (other details as for Figure 1). It is unclear whether [$\alpha$/Fe] trends better with $\Phi$ (panels b and e) or $\Sigma$ (panels c and f), although both show significant improvement on the [$\alpha$/Fe]--M relations (panels a and d).}\label{alpha}
\end{figure*}

\subsection{\texorpdfstring{$\alpha-$}{}Enhancement}
Lastly, Figure~\ref{alpha} shows the results for [$\alpha$/Fe]. Of the three structural parameters investigated, the [$\alpha$/Fe]--$M$ relations are the weakest. The [$\alpha$/Fe]--$M$ trends (Figures~\ref{alpha}a and~\ref{alpha}d) have the lowest Spearman coefficients and highest RMS values. On the other hand, it is unclear whether [$\alpha$/Fe] trends better with $\Phi$ or $\Sigma$. Overall, the [$\alpha$/Fe]--$\Phi$ relation tends to have less residual trend with size compared to [$\alpha$/Fe]--$\Sigma$: $a_r$ = -0.09 and -0.03 for Figures \ref{alpha}b and \ref{alpha}e compared to 0.11 and 0.18 for Figures \ref{alpha}c and \ref{alpha}f. The difference is only marginal, and for the other measures ($\mathrm{RMS_G}$, $\mathrm{RMS_{rm}}$, and $\rho_S$) there is no clear improvement of one over the other. The same is true for the results of the mass-function weighted analysis (see Table \ref{tbl}); the [$\alpha$/Fe]--$\Phi$ relations have slightly lower RMS values, but the strength of $\rho_S$ and the residual trends with radius are the same within the uncertainties. It is clear that both mass \textit{and} size are important in determining [$\alpha$/Fe], however from these results it is not clear whether $\Phi$ or $\Sigma$ better represents this dependence.

\section{Discussion}
For each stellar population (SP) parameter we compared the correlations with each of $M$, $\Phi$, and $\Sigma$ in three ways. Firstly, we use the RMS values, in conjunction with the relative observational uncertainty on the parameters, to understand the relative intrinsic scatter. Secondly, we fit the residuals of the Gaussian model as a function of galaxy size, and use the value of the slope to determine which structural parameter best encapsulates the SP parameter's dependence on size. Thirdly, we use the Spearman correlation coefficient $\rho_S$ as a nonparametric assessment of the strength of the correlations. We find our log-linear relations to be adequate representations of the data, as indicated by the linearity of the running median and the similarity of the RMS values for the Gaussian model and running median fits. Given this, we are able to compare the RMS values for different fits to determine which structural parameter is the best predictor for the stellar population parameter in each case.

From our analysis, we find clear results which show that the SP parameters $g-i$ color and stellar metallicity [Z/H] correlate best with the depth of the gravitational potential $\Phi$, while SP age trends best with surface density $\Sigma$. On the other hand, the results for $[\alpha/Fe]$ are not so clear; the [$\alpha$/Fe]-$\Phi$ relations are only slightly better than [$\alpha$/Fe]-$\Sigma$, although both are appreciably better than the relations with $M$.

\cite{Wake2012}, \cite{Thomas2010} and \cite{Graves2009_1} found that galaxy color and [Z/H] correlate better with $\sigma$ than with either $M_*$ or $M_D$, and the Age-$\Sigma$ relation was explored recently by \cite{Scott2017}. Our analysis builds on these works and others by: (i) quantitatively analyzing residual trends with galaxy size; (ii) comparing the observational uncertainty on the parameters to deduce the relative intrinsic scatter in the relations; and (iii) showing that trends with $\sigma$ are reproduced using the purely photometric estimator for the gravitational potential, $M_*/R_e$.

By understanding the relative intrinsic scatter, we can infer the likelihood of parameters being causally linked. However, without a theoretical framework of the physical processes driving these trends, it remains uncertain whether these correlations represent causation, or are the result of some other underlying trend. We therefore present possible frameworks, while acknowledging that more work is required to determine the true physical impact of these mechanisms in relation to other galactic processes.

\subsection{Sample Selection}
We find very close agreement between the results for the unweighted SAMI sample and the weighted sample representing the galaxy mass function. For $g-i$ color,  [Z/H], and age, the correlations which show the least scatter, least residual trend with size, and highest correlation coefficient in the main analysis are the same as those in the mass-function weighted analysis. The two samples also agree in that $\Phi$ shows only a marginal improvement compared to $\Sigma$ for [$\alpha$/Fe].

\subsection{Color-$\Phi$ Diagram}
Due to the tighter relations in both the RS and BC, we infer that the color-$\Phi$ diagram is a more precise tool than the traditional color-$M$ diagram for identifying a galaxy's evolutionary type. The RS and BC are better aligned in color-$\Phi$ space, indicating a smoother transition between the two populations. Furthermore, the significant residual trend with size in the color--$M$ diagram, indicates galaxy size as well as mass (in the form $M/R_e$) is required to accurately determine observed color.

\subsection{Metallicity-$\Phi$ Relation}
We suggest the stronger correlation between [Z/H] and $\Phi$ (rather than $M$), is evidence that gravitational potential is the main regulator of global SP metallicity. The underlying physical mechanism is that the depth of the gravitational potential determines the escape velocity required for metal-rich gas to be ejected from the system. This hypothesis is supported by the tight radial trend in ETGs between local escape velocity and line strength indices \citep{Scott2009}. Assuming star formation occurs mostly {\it in situ} \citep[e.g.][]{Johansson2012}, we would predict a similar relation using the gas-phase metallicity in star-forming galaxies (D'Eugenio et al. submitted). Even so, we know ETGs have long evolutionary histories that include galaxy mergers, and this hypothesis does not, on its own, explain how the relation is maintained through mergers. However, simulations by \cite{BoylanKolchin_Ma_2007} of the accretion of satellite galaxies found that low-density satellites are easily disrupted, losing a large fraction of their mass during early passes at large radii; high-density satellites are more likely to survive multiple passes and continue sinking towards the center of the host. This maintains the existing [Z/H]--$\Phi$ relation, because diffuse, low--metallicity satellites will lower both the potential and metallicity of the host by adding low--metallicity material at large radii. Conversely, compact, high--metallicity satellites will carry most of their mass into the inner regions of the host, deepening the host's potential and increasing its [Z/H].
\bigskip
\bigskip

\subsection{Age and $\alpha$-Enhancement}
We find strong evidence for the age-$\Sigma$ relation, however it is unclear whether [$\alpha$/Fe] correlates better with $\Phi$ or $\Sigma$; the best correlation may lie somewhere between the two quantities (i.e. [$\alpha$/Fe] $\propto$ $M/R^x$ for $x \in [1,2]$).

Taking [$\alpha$/Fe] as a measure of star formation duration (SFD) and assuming ETGs formed approximately coevally, it follows naturally that a long SFD (low [$\alpha$/Fe]) will correspond to a younger `single-burst' SP; conversely, a short SFD (high [$\alpha$/Fe]) will correspond to an older `single-burst' SP. Thus, if ETGs are coeval, we can expect age and [$\alpha$/Fe] to correlate with the same structural parameter (whichever that may be).

To explain the origin of the correlations with $\Sigma$, we propose the following two mechanisms: (1) compactness-related quenching; and (2) the $\Sigma_{SFR}\propto\Sigma_{gas}$ relation. As we will argue below, both mechanisms appear in broad agreement with our results, although a more detailed semi-analytical approach would help resolve their relative impact on ETG stellar populations.

Quiescence correlates strongly with central surface density, regardless of the measurements used: whether quiescence is measured via specific star formation rate \citep[sSFR;][]{Brinchmann2004,Franx2008,Barro2013,Woo2015,Whitaker2017}, via the fraction of red sequence galaxies \citep[$f_q$;][]{Omand2014}, or some other measure of star formation history \citep[e.g. the $D_n 4000$ break;][]{Kauffmann2003}. \cite{Woo2015} proposed two main quenching pathways which act concurrently but on very different timescales: central compactness-related processes are rapid, while halo quenching is prolonged. Compactness-related processes are those which, as a direct or indirect consequence of building the central bulge, contribute to quenching. For example, gaseous inflows from the disk to the bulge, triggered by disk instability or an event such as a major merger, are exhausted in a star burst, leading to an increased bulge compactness. Furthermore, these inflows can trigger active galactic nuclei, from which the feedback heats and blows away surrounding gas preventing further star formation. In this scenario of compactness-related quenching, it follows that galaxies with a high $\Sigma$ (i.e. compact star formation) quenched faster and hence earlier, resulting in an older SP and a shorter SFD than their diffuse counterparts. This leads naturally to the age-$\Sigma$ and [$\alpha$/Fe]-$\Sigma$ relations in ETGs.

Alternatively, given age--$\Sigma$ and [$\alpha$/Fe]--$\Sigma$, we could look to the $\Sigma_{gas}\propto\Sigma_{SFR}$ relation \citep[e.g.][]{Schmidt1959,Kennicutt1998,Federrath2017} for an empirical explanation. A high $\Sigma_{gas}$ in star-forming disks produces a high specific star formation rate (SFR), and (due to the finite supply of gas) this then leads to a short SFD, and hence an old SP age. This trend with $\Sigma_{gas}$ in the BC becomes fossilized as a trend in $\Sigma_*$ and $\Sigma_D$ in ETGs.

However, neither of these two mechanisms explain why [$\alpha$/Fe] also trends strongly with $\Phi$. A possible interpretation is that the extent to which [$\alpha$/Fe] correlates with $\Phi$ and not $\Sigma$, indicates the extent to which these galaxies are not coeval, and the time since formation as a function of mass and/or size. The residuals of the Gaussian fit in Figures \ref{alpha}c and \ref{alpha}f show that at fixed $\Sigma$, larger galaxies have higher [$\alpha$/Fe], and hence more prolonged star formation histories. Future analyses could focus on analytic or semi-analytic modeling to explain these trends.

\section{Summary}
Our analysis builds on \cite{Franx2008} and \cite{Wake2012}, arguing that the evolution of stellar populations is driven by physical parameters other than galaxy mass. We find the tightest correlations, and the least residual trend with galaxy size, for the $g-i$ color--$\Phi$, [Z/H]--$\Phi$, and age--$\Sigma$ relations. We find [$\alpha$/Fe] to correlate strongly with both $\Sigma$ and $\Phi$. We show that correlations with $\sigma$ are reproduced using the purely photometric $M_*/R_e$. From these results, our inferences and interpretations are that: (1) the color--$\Phi$ diagram is a more precise tool for determining the developmental stage of the stellar population than the color--mass diagram; (2) gravitational potential is the primary regulator for global stellar metallicity, via its relation to the gas escape velocity.  We also propose two possible mechanisms for the age--$\Sigma$ and [$\alpha$/Fe]--$\Sigma$ correlations: the age--$\Sigma$ and [$\alpha$/Fe]--$\Sigma$ correlations are results of compactness-driven quenching mechanisms; and/or the correlations are fossil records of the $\Sigma_{SFR}\propto\Sigma_{gas}$ relation in their disk-dominated progenitors. Determining which of the various possible physical mechanisms are responsible for these relations requires comparison to detailed simulations that take into account of all these processes.

\acknowledgments
The SAMI Galaxy Survey is based on observations made at the Anglo-Australian Telescope. The SAMI spectrograph was developed jointly by the University of Sydney and the Australian Astronomical Observatory. The SAMI input catalog is based on data from the Sloan Digital Sky Survey, the GAMA Survey and the VST ATLAS Survey. The SAMI Galaxy Survey is funded by the Australian Research Council Centre of Excellence for All-sky Astrophysics (CAASTRO; grant CE110001020), and other participating institutions.

TMB is supported by an Australian Government Research Training Program Scholarship. NS acknowledges a University of Sydney Postdoctoral Research Fellowship. JvdS is funded under JBH's ARC Laureate Fellowship (FL140100278). SB and MSO acknowledge Australian Research Council Future Fellowships (FT140101166 and FT140100255). AMM acknowledges NASA Hubble Fellowship (HST-HF2-51377) from the Space Telescope Science Institute, operated by Association of Universities for Research in Astronomy, Inc., for NASA (NAS5-26555).

We make extensive use of the Python programming language, including packages SciPy \citep{SciPy}, Astropy \citep{Astropy}, matplotlib \citep{Hunter2007}, emcee \citep{emcee}, and Pathos \citep{Pathos1,Pathos2}. In preliminary analyses we also used TOPCAT \citep{Taylor2005}.

\bibliographystyle{aa}

\end{document}